\documentstyle{article}
\begin{document}

\title{ \bf 
Gravity Quantized
\\ {\Large {\it \ \ (the high tension string) } }
}

\author{
S. Bellucci \\
INFN-Laboratori Nazionali di Frascati, P.O. Box 13,
00044 Frascati, Italy\thanks{bellucci@lnf.infn.it}
\\ \\
and
\\ \\
A. Shiekh\thanks{Presented by A. Shiekh at the 
XX International Workshop  on the Fundamental Problems of High
Energy Physics and Field Theory, Protvino, Russia, June 1997.} \\
International Centre for Theoretical Physics,
Strada Costiera 11, Trieste, Italy\thanks{shiekh@ictp.trieste.it}
}

\date{}

\maketitle

\vskip -8cm
\rightline{LNF-97/012 (P)}
\vskip  8cm

\begin{abstract}
A candidate theory of gravity quantized is reviewed.
\end{abstract}

\baselineskip = 18pt

\section{The perturbative approach}

\subsection{Minimal gravity}

This short review following and extending on a recent proposal
[Shiekh, 1994, 1995, 1996, 1997; Akhundov, Bellucci and
Shiekh, 1996; Bellucci and Shiekh; 1997] is about Einstein gravity
quantized. Why the strange wording? It is well known that
Einstein gravity is {\it not} quantizable; this however does not
preclude the existence of a quantum form, and this talk is all
about this subtle but important difference.

It is well known that Einstein gravity fails to quantize
for the simple reason that the infinities cannot be
accommodated within the starting Lagrangian. For the
purpose of illustration we will be discussing minimal
coupled gravity with massive scalar particles, as
governed by the Lagrangian:

\begin{equation}
{\mathcal L} = \sqrt {-g}
\left( {-2 \Lambda + R + {\textstyle{1 \over 2}}
g^{\mu \nu} ( \partial_\mu \phi )( \partial_\nu \phi ) +
{\textstyle{1 \over 2}} m^2 \phi^2}
\right)
\label{eq:starting}
\end{equation}
\\
The counter terms that carry the infinities cannot be
accommodated back within this starting Lagrangian, and
so the theory retains its divergent nature.

One often speaks of the starting Lagrangian as the
classical Lagrangian arguing that this is the starting
point for quantization, while the final Lagrangian,
which is of the same form, is referred to as the
quantized Lagrangian. This is a misleading notation,
as the original Lagrangian is divergent, having taken
up the counter terms, and the classical limit actually
arises in the $\hbar \to 0$ limit from the final complete
Lagrangian. This distinction will become especially poignant
in what follows.

It is worth repeating for clarity that Einstein gravity
is not quantizable, and we will be making no attempt to
get around this fact.

\subsection{Maximal gravity}

Having noted that the failure to quantize minimal gravity
stemmed from the fact that the counter terms did not fall
back into the starting Lagrangian, one can resort to
extending the Lagrangian so as to ensure that the theory
is `formally' renormalizable. In this way we arrive at
maximal gravity, which is constrained by symmetry to be:

\begin{equation} 
{\mathcal L}_0 = \sqrt{-g_0} 
\left(\matrix { -2\Lambda_0 + R_0 + 
\textstyle{1 \over 2} p_0^2 + 
\textstyle{1 \over 2} m_0^2 \phi_0^2 +
\textstyle{1 \over 4!} \phi_0^4
\lambda_0(\phi_0^2) + p_0^2 \phi_0^2
\kappa_0(\phi_0^2) + R_0 \phi_0^2 \gamma_0(\phi_0^2) \cr
\cr + p_0^4 a_0 (p_0^2,\phi_0^2) + R_0 p_0^2
b_0(p_0^2,\phi_0^2) + R_0^2 c_0(p_0^2,\phi_0^2) + R_{0\mu\nu}
R_0^{\mu\nu} d_0(p_0^2,\phi_0^2) + ... }
\right)
\label{eq:maximal}
\end{equation}

\rightline{ \small \it 
(using units where $16 \pi G = 1$, $c = 1$)}

\noindent where $p_0^2$ is shorthand for $g_0^{\mu \nu}
(\partial_\mu \phi_0) (\partial_\nu \phi_0)$ and not the
independent variable of Hamiltonian mechanics.
$\lambda_0$, $\kappa_0$, $\gamma_0$, $a_0$,
$b_0$, $c_0$, $d_0$ ... are arbitrary analytic functions, and
the second line carries all the higher derivative terms.

Strictly this is formal in having neglected gauge fixing and
the resulting presence of ghost particles. Quantum anomalies
arise from a conflict between symmetries, where only one can be
maintained [Mann, 1988]. For this reason no such trouble is
present here. Had we had massless particles present, we would
accept the conformal anomaly as disrupting the conformal
symmetry.

The price for having achieved `formal' renormalization, is
that the theory (with its infinite number of arbitrary
renormalized parameters) is now devoid of predictive content,
even if it is finally finite.
The failure to quantize has been rephrased from a problem
of non-renormalizability to one of non-predictability.

Despite this, after renormalization we are lead to:

\begin{equation}
{\mathcal L} = \sqrt{-g} 
\left(\matrix { -2\Lambda + R + 
\textstyle{1 \over 2} p^2 + 
\textstyle{1 \over 2} m^2 \phi^2 +
\textstyle{1 \over 4!} \phi^4 \lambda(\phi^2) + p^2 \phi^2
\kappa(\phi^2) + R \phi^2 \gamma(\phi^2) \cr \cr + p^4 a
(p^2,\phi^2) + R p^2 b(p^2,\phi^2) + R^2 c(p^2,\phi^2) +
R_{\mu\nu} R^{\mu\nu} d(p^2,\phi^2) + ... }
\right)
\label{eq:renormalized}
\end{equation}

\subsection{Physical criteria}

Up to this point we have just rephrased the problem
of the non-renormalizability of gravity. Again we take
the liberty to emphasize that we are no longer quantizing
Einstein gravity, but rather some hideously large theory
under the name maximal gravity.

However, there remain physical criteria to pin down some of
these arbitrary factors. Since in general the higher
derivative terms lead to acausal classical behavior, their
renormalized coefficient can be put down to zero on physical
grounds. This still leaves the three arbitrary functions:
$\lambda(\phi^2)$, $\kappa(\phi^2)$ and
$\gamma(\phi^2)$, associated with the terms $\phi^4$,
$p^2 \phi^2$, and
$R \phi^2$ respectively. The last may be abandoned on the
grounds of defying the equivalence principle. To see this,
begin by considering the first term of the Taylor expansion,
namely $R\phi^2$; this has the form of a mass term and so one
would be able to make local measurements of mass to determine
the curvature, and so contradict the equivalence principle
(charged particles, with their non-local fields have this term
present with a fixed coefficient). The same line of reasoning
applies to the remaining terms, $R\phi^4$, $R\phi^6$, ... etc.

This leaves us the two remaining infinite families of
ambiguities with the terms $\phi^4\lambda(\phi^2)$ and
$p^2\phi^2\kappa(\phi^2)$. In the limit of flat space in 3+1
dimensions this will reduce to a renormalized theory in the
traditional sense if $\lambda(\phi^2)=constant$, and
$\kappa(\phi^2)=0$. So one is lead to proposing that the
physical parameters should be:

\begin{equation}
\matrix {
\Lambda = \kappa(\phi^2) = \gamma(\phi^2) = 0 \cr \cr
a(p^2,\phi^2) = b(p^2,\phi^2) = c(p^2,\phi^2) = d(p^2,\phi^2)
=... =0 \cr \cr
\lambda(\phi^2) = \lambda = {\it scalar\ particle\ self\
coupling\ constant} \cr \cr m = {\it mass\ of\ the\ scalar\
particle} }
\end{equation}
\\
\noindent and so the renormalized theory of quantum gravity
for a scalar field should have the form:

\begin{equation} 
{\mathcal L} = \sqrt{-g} 
\left( R + \textstyle{1 \over 2} 
g^{\mu \nu} ( \partial_\mu \phi )( \partial_\nu \phi ) + 
\textstyle{1 \over 2} m^2 \phi^2 +
\textstyle{1 \over 4!} \lambda \phi^4
\right)
\label{eq:final}
\end{equation}
\\
This is a candidate for the long sought after Einstein gravity
quantized; and not quantized Einstein gravity, and the classical
theory arises in the $\hbar \to 0$ limit of this.

\section{Self consistency}

One might now worry about the renormalization group pulling
the coupling constants around.

Since we are interested only that the zeroed couplings
remain so, we shall name them as external couplings,
in so much as they belong to terms outside the final
renormalised Lagrangian (eq.~\ref{eq:final}).
The finite number remaining  will naturally take up the
designation of internal couplings.

When a coupling runs, its value at some scale must be
specified. The beta function then determines how the
coupling varies for other scales. It can now be seen
to be a trivial matter to stop the external couplings
from running, namely by zeroing them at an infinite scale.

\section{Non-perturbative perspective}

The above argument was done completely within a perturbative
context, and one might wonder if a non-perturbative perspective
would lead to the same proposal, and then perhaps without the
infinities of the perturbative approach.

\subsection{The high tension string}

String theory might be thought of as another attempt
to quantize gravity by generalizing away from point
particle theory (super-gravity having failed).

When viewing string theory as a higher derivative,
infinitely large Lagrangian, one sees many similarities
with orthodox gravity, excepting that string theory
has only one, and not an infinity, of extra parameters
in the form of the string tension.

It then becomes very natural to wonder about the
point particle limit of the super-string, when one
anticipates the appearance of super-gravity. There
immediately arises a question of how to resolve
the fact that super-gravity is not renormalizable, but
that the string in the high tension limit exists.
The above investigation makes the resolution rather
transparent in so much as the starting Lagrangian
is not that of super-gravity even in the limit,
for one has no reason to suppose the higher derivative
bare terms disappear in the high tension limit.
Again, only the theory quantized reduces to super-gravity.

In this way one might view orthodox gravity as the
point particle (high tension) limit of the string,
and as such it is a second confirmation of the existence
of orthodox gravity as a candidate for gravity quantized.

\subsection{Occam's gravity}

We generalised to super-gravity and string theory because
our former candidates failed to give us quantum gravity.
But why go to the complexities of higher dimensions, or
a new set of particles if we can locate a simpler candidate?
Naturally, at the end of the day, it is not a choice for
us to make, but rather a question to be put to nature.

It is rather paradoxical that we have arrived at a minimalist
proposal by having first resorted to a maximal theory.
But it is satisfying in having added every ingredient to
the broth and seeing it slim itself down on its
own accord.

Being such a simple candidate, one can immediately go about
calculating with this proposal. Other benefits of the
simplicity are that setting the external couplings leads
to no further Feynman rules than for normal gravity, and
implies the use of minimal subtraction for the gravitational
infinities that fall outside of the inner Lagrangian.

\section*{Acknowledgments}
This work does not reflect the views of the High Energy
Physics group at ICTP.

\section*{References}

\begin{description} {\small

\item[$\bullet$] \hskip .54cm {\bf A. Shiekh}, 
{\it `The Perturbative Quantization of Gravity'}, in 
``Problems on High Energy Physics and Field Theory'',
Proceedings of the XVII workshop 1994, pp. 156-165,
Protvino, Russia, 1995, {\tt hep-th/9407159}.
\\
{\bf A. Shiekh}, 
{\it `Can the Equivalence Principle Survive Quantization?'},
in the ``International Workshop on Anti-Matter and Anti-Hydrogen
Spectroscopy'', Molise, Italy, 1996, {\tt gr-gc/9606007}.
\\
{\bf A. Shiekh}, 
{\it `Is there no quantum form of Einstein Gravity?'}, in 
``Problems on High Energy Physics and Field Theory'',
Proceedings of the XIX International workshop 1996,
Protvino, 1997, {\tt gr-qc/9607005}.
\\
{\bf A. Shiekh}, 
{\it `Quantizing Orthodox Gravity'}, Can. J. Phys., 
{\bf 74}, 1996, 172-, {\tt hep-th/9307100}.
\\
{\bf S. Bellucci, A. Shiekh}, 
{\it `Consistency of Orthodox Gravity'}, 
{\tt gr-qc/9701065}, LNF-97/003 (P).
\\
{\bf A. Akhundov, S. Bellucci, A. Shiekh},  {\it `Gravitational
interaction to one loop in effective quantum gravity'}, 
{\tt gr-qc/9611018}, LNF-96/058 (P), Phys. Lett. B, {\bf 395}, 16-.
\\
\item[$\bullet$] \hskip .54cm {\bf R. Mann}, {\it `Zeta
function regularization of Quantum Gravity'}, In Proceedings
of the cap-nserc Summer Workshop on Field Theory and Critical
Phenomena. Edited by G. Kunstatter, H. Lee, F. Khanna and H.
Limezawa, World Scientific Pub. Co. Ltd., Singapore, 1988, p.
17-. 
\\
\item[$\bullet$] \hskip .54cm {\bf P. Ramond}, {\it ``Field
Theory: A Modern Primer''}, 2nd Ed (Addison-Wesley, 1990).

}
\end{description}

\end{document}